\documentclass{article}

\usepackage{arxiv}

\usepackage[utf8]{inputenc} 
\usepackage[T1]{fontenc}    
\usepackage{hyperref}       
\usepackage{url}            
\usepackage{booktabs}       
\usepackage{nicefrac}       
\usepackage{microtype}      
\usepackage{lipsum}
\usepackage{graphicx}
\usepackage{rotating}
\usepackage{multirow}%
\usepackage{amsmath,amssymb,amsfonts}%
\usepackage{amsthm}%
\usepackage{mathrsfs}%
\usepackage[title]{appendix}%
\usepackage{textcomp}%
\usepackage{manyfoot}%
\usepackage{algorithm}%
\usepackage{algorithmicx}%
\usepackage{algpseudocode}%
\usepackage{listings}%

\usepackage{multirow}
\usepackage{adjustbox}
\usepackage{subcaption}
\usepackage{graphicx}     
\usepackage{tabularx}
\usepackage{array}
\usepackage{booktabs}
\graphicspath{ {./images/} }

\title{Training chord recognition models on artificially generated audio}

\author{
 Martyna Majchrzak \\
  Faculty of Mathematics and Information Science\\
  Warsaw University of Technology\\
  Koszykowa 75, 00-662 Warsaw, Poland \vspace{0.5em} \\
  \texttt{martyna.majchrzak19@gmail.com} \\
   \And
 Jacek Ma{\'n}dziuk \\
  Faculty of Mathematics and Information Science\\
  Warsaw University of Technology\\
  Koszykowa 75, 00-662 Warsaw, Poland \vspace{0.5em} \\
  \texttt{mandziuk@mini.pw.edu.pl} \\
}

\begin{document}
\maketitle
\begin{abstract}
One of the challenging problems in Music Information Retrieval is the acquisition of enough non-copyrighted audio recordings for model training and evaluation. This study compares two Transformer-based neural network models for chord sequence recognition in audio recordings and examines the effectiveness of using an artificially generated dataset for this purpose. The models are trained on various combinations of Artificial Audio Multitracks (AAM), Schubert's Winterreise Dataset, and the McGill Billboard Dataset and evaluated with three metrics: Root, MajMin and Chord Content Metric (CCM). The experiments prove that even though there are certainly differences in complexity and structure between artificially generated and human-composed music, the former can be useful in certain scenarios. Specifically, AAM can enrich a smaller training dataset of music composed by a human or can even be used as a standalone training set for a model that predicts chord sequences in pop music, if no other data is available.
\end{abstract}

\bigskip
\bigskip

\section{Introduction}
\label{sec:introduction}

Music Information Retrieval (MIR) is
an interdisciplinary area of research that involves fields such as musicology, signal processing, informatics, and machine learning. MIR encompasses various aspects of music-related activities, such as \textit{music classification} (genres and composers)~\cite{classification:deep,classification:combined,classification:an:evolutionary}, \textit{music recommending systems}~\cite{Tran2023Emotion,Song2012SurveyRecommendation,recommendation:a:semipersonalized}, \textit{melody harmonization}~\cite{MYCKA2023TowardHumanLevel,Mycka2022HumanLevelMelodicLineHarmonization,Mycka2022ICAISC}, \textit{music composition}~\cite{wang2024review,mandziuk2014neuro,mandziuk2013chopin}, \textit{music transcription}~\cite{sturm2016music}, and many others~\cite{mycka2025artificial}.

One of the central tasks within MIR is Automatic Chord Recognition (ACR), which consists in dividing an audio recording or a sequence of features extracted from such a recording into segments and labeling each segment with a name of a musical chord present in that segment. 

\subsection{Motivation.}
ACR is a complex task, and the collection of reliable reference data is one of the main impediments and challenges in the application of deep neural networks or other machine learning (ML) models to solve this task.

Manual data annotation is tedious and time-intensive, and even skilled musicians may disagree on labeling certain musical fragments. Moreover, many open source datasets only share annotations, and not audio files, due to copyright issues. In effect, suitable ready-to-use datasets are scarce and generally insufficient for training complex ML models. Recent advances in ML, including the development of music generation systems, have expanded these possibilities. However, a question arises regarding \textbf{the efficacy of models} trained on artificially generated (composed) music when applied to music pieces composed by humans. This fundamental question lies at the heart of the research presented in this paper.

\subsection{Contribution.}
The main contribution of this work is fourfold:
\begin{itemize}
    \item analysis of structural annotation differences in considered ACR datasets,
    \item systematization of metrics used for ACR model evaluation,
    \item performance comparison of ACR models trained on various combinations of artificial and human-composed music,
    \item guiding remarks regarding the potential effect of including artificially-generated music into a training set composed of human-generated data.
\end{itemize}

\section{Related work}
\label{sec:related-work}

First chord recognition systems were mostly knowledge-based, specifically template-based~\cite{Fujishima1999RealtimeCR,templatebased2009} and HMM-based~\cite{HMM2003,HMM2006,4275055,HMM2008}. With the development of data-driven methods, the focus has been shifted toward the utilization of neural networks for this task. A neural network can be used either as a tool to create chroma features from audio recordings, or as a classifier based on already created features, or as an end-to-end solution. One of the first works utilizing feedforward neural networks for Chord Recognition was~\cite{neural_networks_2012}, which applies 12-valued Pitch Class Profiles and a simple neural architecture with 35 neurons in a hidden layer. The model was evaluated on a self-created database of chords (mostly guitar ones). Subsequent publications used neural networks comprised of rectifier linear units~\cite{DeepChromaExtractor2016,koops2017chord}, Convolutional Neural Networks~\cite{6406762,DeepLearning2015,DBLP:journals/corr/KorzeniowskiW16a}, Recurrent Neural Networks~\cite{RecurrentNeuralNetworks,8282235}, hybrid Convolutionally Recurrent Neural Networks~\cite{CRNN_MIREX2018}, Variational Auto Encoders~\cite{DBLP:journals/corr/abs-2005-07091} and, more recently, Transformers~\cite{Chen2019HarmonyTI,DBLP:journals/corr/abs-1907-02698}. Some studies \cite{doi:10.1080/09298215.2019.1613436},  \cite{Pauwels201920YO} confirmed that annotator subjectivity is an important factor for chord recognition systems, but modern algorithms are powerful enough to tune themselves to the personal factors influencing particular annotators' decisions.

\subsection{Transformer architecture for ACR}
With the growing popularity of the Transformer architecture \cite{DBLP:journals/corr/VaswaniSPUJGKP17}, originally developed for Natural Language Processing asks, some researchers proposed its application for chord recognition. A Harmony Transformer~\cite{Chen2019HarmonyTI} assignes the chord segmentation task to the encoder and the chord recognition task to the decoder. Another work introduces the Bidirectional Transformer for Chord Recognition~\cite{DBLP:journals/corr/abs-1907-02698}, highlights the usefulness of the attention mechanism, and visualizes the way the model works through attention maps. One of the advantages of the transformer architecture is the the lack of the requirement for additional decoders such as HMMs or Conditional Random Fields (CRFs), so only one training phase is required.

\subsection{Datasets}
The first common dataset made available to researchers was released by Harte et al. in 2005 \cite{Harte2005}. It consists of chord annotations for twelve studio albums by the Beatles and was later released as one of the Isophonics Datasets \cite{Isophonics} (other datasets include musical pieces by Zweieck, Queen and Carole King). Subsequent years brought about efforts to create larger and more diverse datasets. The majority of them include annotations and some kind of chroma features, and not actual audio files. The Billboard annotations dataset \cite{mcgill_billboard} covers a broad range of artists and musical genres, and includes NNLS (Non-negative Least Squares) chroma vectors~\cite{NNLS} and tuning estimates from the Chordino VAMP plugin \cite{NNLSChroma}.
The RWC Music Database\cite{RWC} publishes MFCC (Mel-Frequency Cepstral Coefficients) features, but is not released in an open-acess mode. 
The Schubert Winterreise Dataset \cite{Schubert_Winterreise} is an exception in that regard - it includes not only audio recordings, but also a variety of metadata, including chord annotations. It is, however, a very small sample of classical music. Recently, Artificial Audio Multitracks \cite{AAM2023}, a selection of 3000 artificial music tracks created by an algorithmic composer, was released.

\subsection{Metrics}
The question of how to evaluate chord recognition systems results does not only refer to the selection of a suitable metric, but also concerns the selection of a set of labels (referred to as a vocabulary) and a comparison strategy. The metrics can be defined on a single chord level or on a chord sequence level. 

A \textbf{single chord level} metric simply compares two chord labels. Examples of this type of metric include
\begin{itemize}
    \item \textbf{Root} - an indicator of whether the roots of the compared chords are the same. For example, C:maj and C:min chords have the same root note, so root(C:maj, C:min)=1. 
    \item \textbf{MajMin} - compares major, minor, and “no chord” labels and indicates whether they are the same. If the vocabulary contains more complicated chords, they are simplified and assigned to one of the 25 classes and then compared.
    \item \textbf{Mirex} - an estimated chord is considered correct if it has at least three pitch classes in common with the ground truth annotation.
    \item \textbf{Chord Content Metric (CCM)}
\cite{DBLP:journals/corr/abs-2201-05244} accounts for (possibly overlapping) notes that the predicted and reference chords have in common. Unlike the previously defined ones, it's not a binary metric and takes values between 0 and 1, according to the following definition:
$$ A = \frac{C - I + |y|}{2 |y| }$$
Where $y$ is the number of notes in the annotation chord, $\hat{y}$ - number of notes in the predicted chord, $C = |y \cap \hat{y}|$ - number of correctly identified notes, and $ I = | \hat{y} \ y|$ - number of extra predicted notes that were present in the annotation.
\end{itemize}
A \textbf{chord sequence level} metric compares two sequences of chords, containing information about the durations of each labeled sequence piece. Typical examples include:
\begin{itemize}
\item \textbf{WCSR \textbackslash WAOR} -
Weighted Chord Symbol Recall \textbackslash Weighted Average Overlap Ratio - a total duration of segments with a correct prediction.
\item \textbf{Weighted Accuracy} - a generalization of WCSR, suitable for non-binary metrics, such as CCM. It is the average value of the desired metric for all chords in the sequence, weighted by the duration of each segment.
\end{itemize}

\section{Proposed approach}\label{sec:approach}

Datasets for Audio Chord Recognition that contain actual audio recordings are limited in size and accessibility. Artificially generating them could expand the volume of potential training data, but a question arises of how helpful are systems trained on those datasets for recognizing chords in audio recordings that was not artificially generated.

To address the above question, in this study, the AAM dataset is used as an example of an artificially generated audio dataset that includes chord annotations, and the Billboard and Winterreise datasets as examples of two genres of human-composed music: popular and classical. Two model architectures from the recent literature are trained and evaluated on different combinations of those datasets to examine if any useful conclusions can be drawn from differences in their performance. Furthermore, to increase the generality of the conclusions one model uses raw audio as its input and the other one takes chroma features as input.

\subsection{Model architectures}
Both based are based on the transformer architecture: the first one is Bi-directional Transformer for chord recognition (BTC) \cite{DBLP:journals/corr/abs-1907-02698} and the second one is Harmony Transformer (HT) \cite{Chen2019HarmonyTI}. They were chosen primarily due to being proposed fairly recently (2019) and the availability of the implementation code. They both incorporate the information from the surrounding frames into the prediction process, so they predict a sequence of chords instead of a single chord.

\subsection{Datasets.}

\subsubsection{The McGill Billboard Project} \label{The McGill Billboard Project}
The Billboard annotations dataset~\cite{mcgill_billboard} covers a broad range of artists and musical genres. The songs were sampled from the \textit{Billboard} ``Hot 100'', a weekly compilation of the most popular music in the USA. The audio tracks are not included in the datasets due to copyright restrictions. However, the authors were able to include audio features: NNLS chroma vectors~\cite{NNLS} and tuning estimates from the Chordino VAMP plugin~\cite{Chordino_WAMP}.
Original, complete annotations include the chords, song structure, instrumentation, and timing in a format resembling musical scores. However, the authors also included the MIREX style .lab files, with start time, end time, and chord labels. One can download the files with two different chord-label dictionaries - a more extensive one and a simplified one. This dataset is publicly available online~\cite{McGill_Billboard_Project} 
and has been used in multiple studies, such as~\cite{RecurrentNeuralNetworks}, \cite{Sigtia2015AudioCR},  \cite{Humphrey2015FourTI}, \cite{korzeniowski2018improved} \cite{Chen2019HarmonyTI}, \cite{CRNN_MIREX2018}, \cite{CRNN_MIREX2019} and \cite{8902741}.

\subsubsection{Schubert Winterreise} \label{Schubert Winterreise}
The Schubert Winterreise dataset is a recently introduced multimodal dataset~\cite{Schubert_Winterreise} that consists of Franz Schubert’s 24-song cycle called 'Winterreise', composed in 1827 for voice and piano. One of its unique features is its availability in several representations: audio .wav files, lyrics (in German) as well as scores in different formats, such as midi and PDF. In addition, it contains chord and note annotations for all the songs (both the recording of nine different performances and the scores), with start and end times (in seconds) available. Out of those nine performances, only two are actually included in the audio form due to copyright issues. The length of the raw audio data is 2 hours, 14 minutes and 16 seconds (1:07:31 for one of the performances and 1:06:45 for the other). Each chord label is provided in 4 notations: shorthand, extended with explicit intervals, reduced to major and minor triads, and reduced to major and minor triads with a bass note. Moreover, the dataset contains annotations for the audio structure (repetitions of parts of the song), global and local keys, as well as additional metadata and files used for the preprocessing of the recordings.
The dataset is available online~\cite{Schubert_Winterreise_1}
under a Creative Commons Attribution 3.0 Unported license. Althought the dataset is relatively new it has already proven useful for a variety of different studies, including local key estimation~\cite{9054642} and learning pitch-class representations \cite{Pitchclass2021} \cite{DeepPitchclass2021}.

\subsubsection{Artifcial Audio Multitracks} \label{Artifcial Audio Multitracks}
Artificial Audio Multitracks \cite{AAM2023} is a state-of-the-art dataset of 3000 artificial music tracks with rich annotations based on real instrument samples generated by algorithmic composition with respect to music theory. It is intended for various music information retrieval tasks like music segmentation, instrument recognition, source separation, onset detection, key, and chord recognition. As the audio is perfectly aligned to the original MIDIs, all annotations (onsets, pitches, instruments, keys, tempos, chords, beats, and segment boundaries) are absolutely precise. Furthermore, the authors conducted experiments proving that this dataset is useful for neural network models for music segmentation, instrument recognition, and onset detection. However, no such study has yet been done for chord recognition. For each track, MIDI, mp3 files (as separate instruments and mixes) and annotations are available. The entire dataset is available online~\cite{Artificial_Audio_Multitracks}. 

\subsection{Exploratory Data Analysis}

%


\begin{figure}[!h]
  \centering
  \begin{subfigure}{0.49\textwidth}
    \centering
    \includegraphics[width=\textwidth]{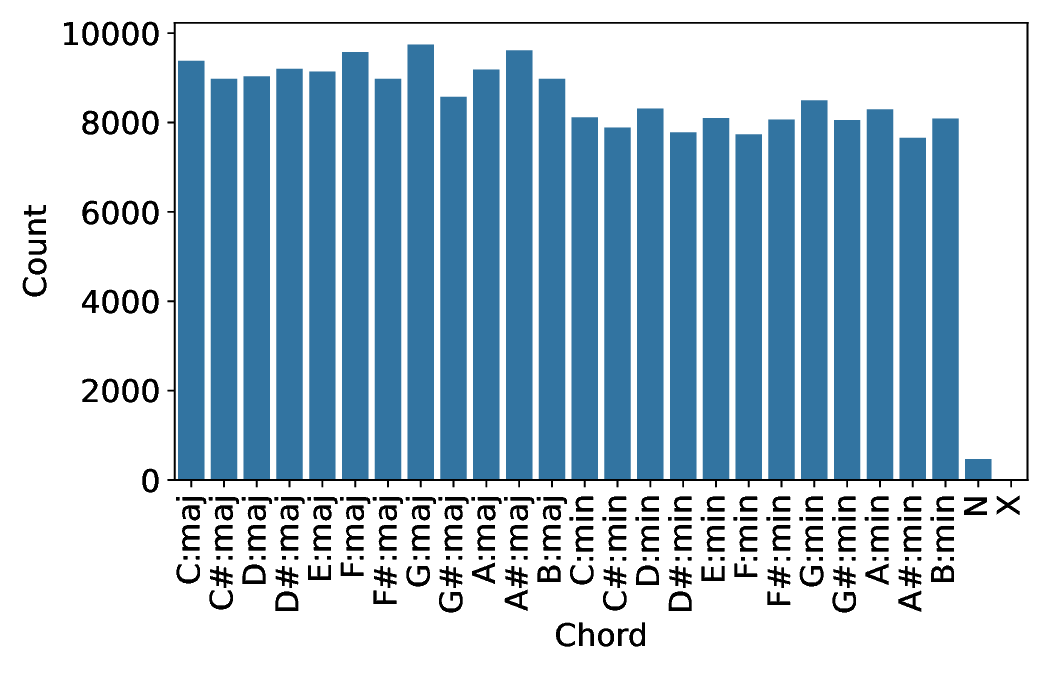}
    \caption{Chord occurrences in AAM dataset}
    \label{fig_co_aam}
  \end{subfigure}%
  \hfill
  \begin{subfigure}{0.49\textwidth}
    \centering
    \includegraphics[width=\textwidth]{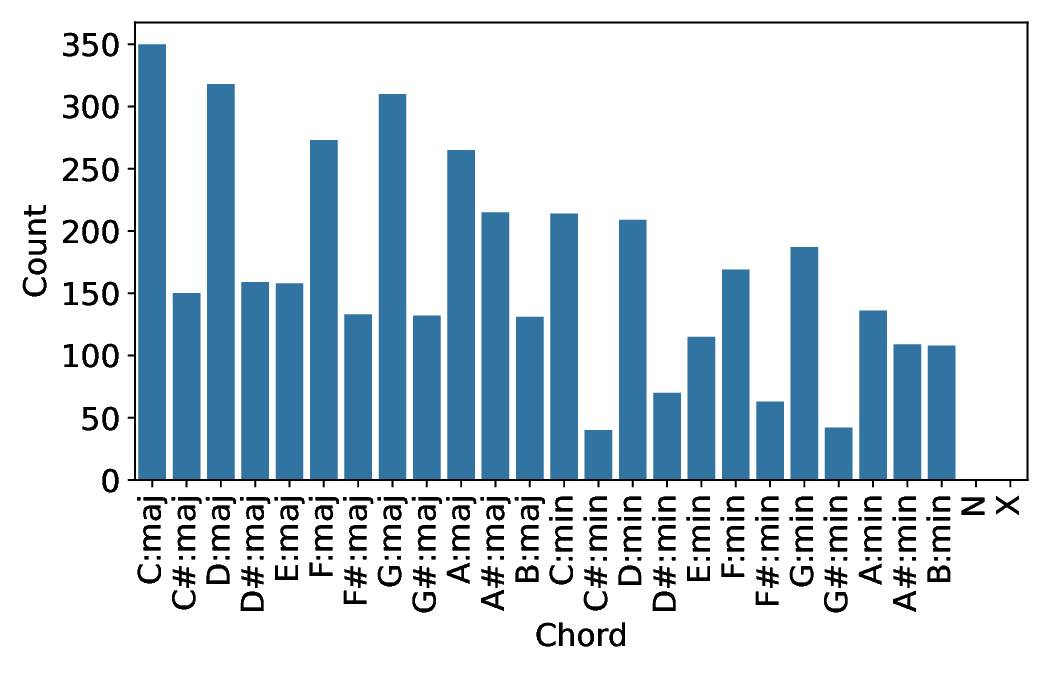}
    \caption{Chord occurrences in Winterreise dataset}
    \label{fig_co_winterreise}
  \end{subfigure}

  \vspace{1em} 

  \begin{subfigure}{0.49\textwidth}
    \centering
    \includegraphics[width=\textwidth]{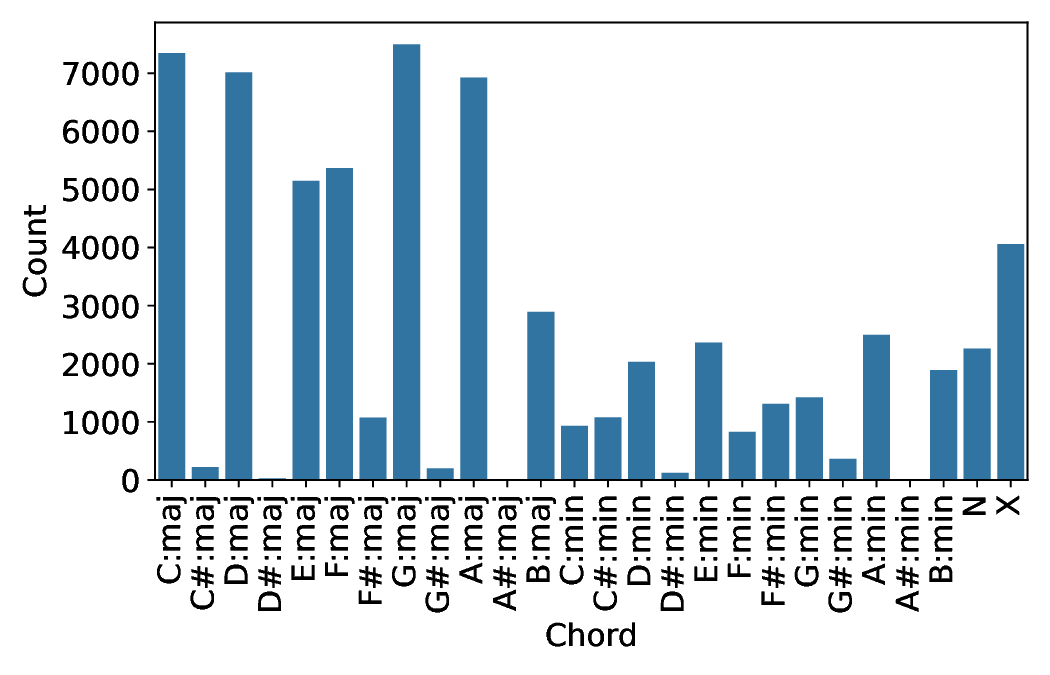}
    \caption{Chord occurrences in Billboard dataset}
    \label{fig_co_billboard}
  \end{subfigure}
  \caption{Number of occurrences of each chord from the vocabulary in the AAM, Winterreise, and Billboard datasets. If the chord label in the annotation file is present for several consecutive rows, it is counted once (as one occurrence).}
  \label{fig_co}
\end{figure}

To explore the differences 
of the annotation
structure in the considered datasets, Figure~\ref{fig_co} presents the per-dataset frequency of each of the Major and Minor chords in.

It is clearly visible in the figure \ref{fig_co} 
that the distribution of chords in the AAM dataset 
is much more uniform than in the datasets that consist of human-composed songs. In 
the Billboard dataset, some chords, such as A\#:maj and A\#:min are never present. The low number of Non-chord labels in AAM and Winterreise compared to Billboard, is caused by a different way the annotation files are constructed. For the first two datasets the annotations start at the moment where a chord is present and recognized, and for the latter one, they start right at the second 0.00 and the silent interval at the beginning and at the end of the file are annotated as 'N'.

\begin{figure}[!h]
  \centering
  \begin{subfigure}{0.5\textwidth}
    \centering
    \includegraphics[width=\textwidth]{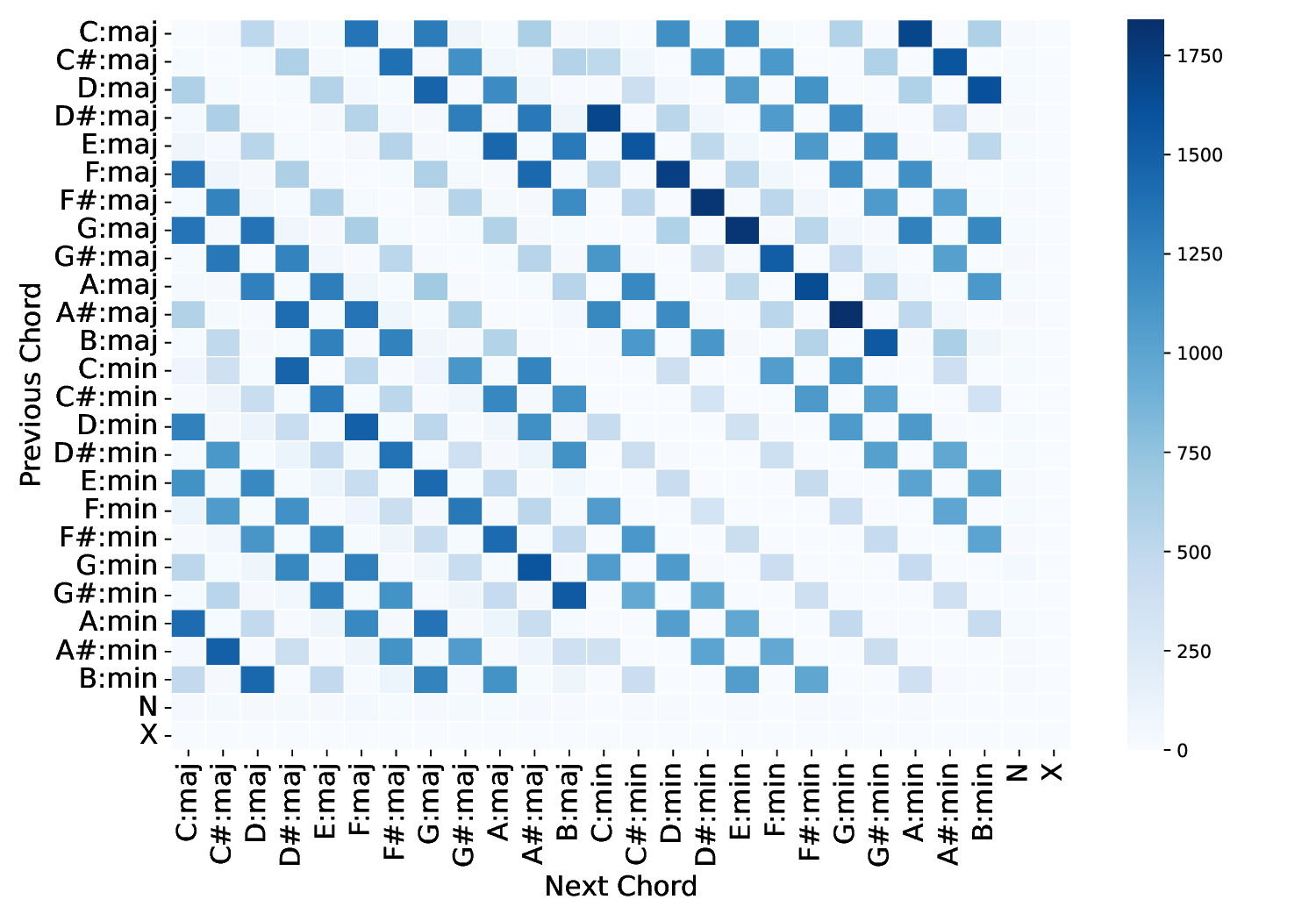}
    \caption{Chord progressions in AAM dataset}
    \label{fig_cp_aam}
  \end{subfigure}%
  \hfill
  \begin{subfigure}{0.5\textwidth}
    \centering
    \includegraphics[width=\textwidth]{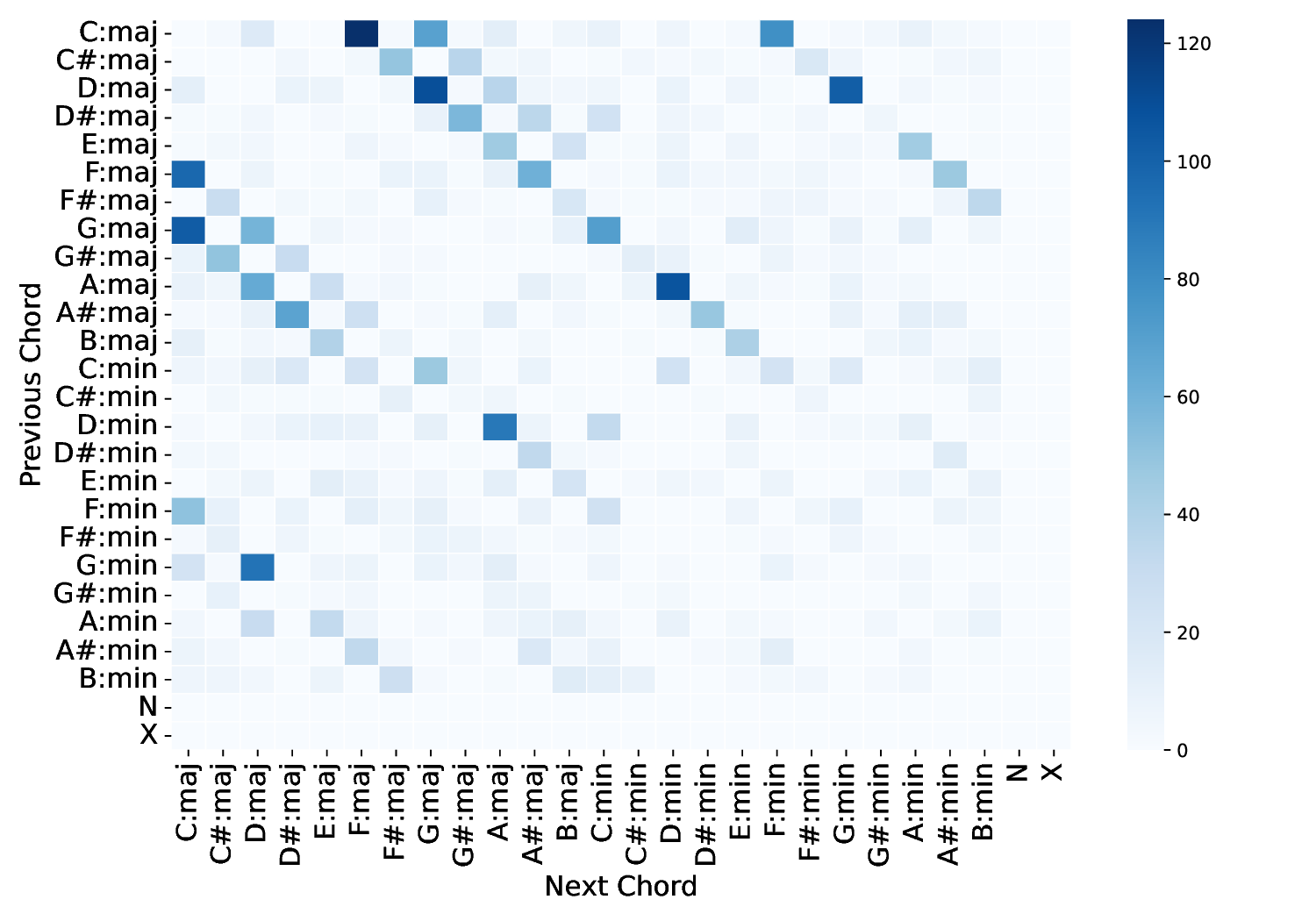}
    \caption{Chord progressions in Winterreise dataset}
    \label{fig_cp_winterreise}
  \end{subfigure}

  \vspace{1em} 

  \begin{subfigure}{0.5\textwidth}
    \centering
    \includegraphics[width=\textwidth]{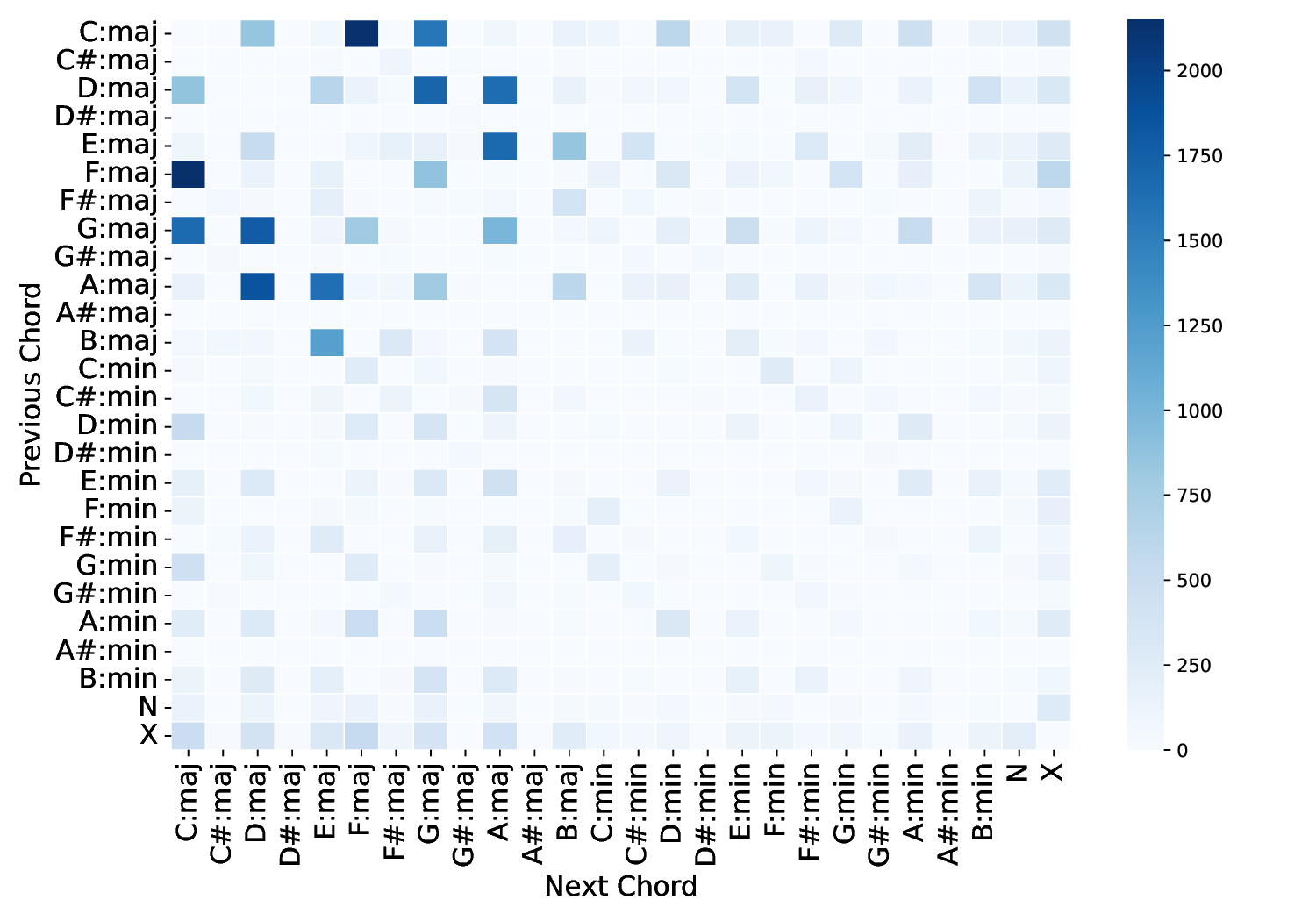}
    \caption{Chord progressions in Billboard dataset}
    \label{fig_cp_billboard}
  \end{subfigure}
  \caption{Number of changes from one chord to another in the AAM, Winterreise and Billboard datasets. Only~changes to a different chord are counted.}
  \label{fig_cp}
\end{figure}

Figure \ref{fig_cp} show counts of each possible chord progression in each of the datasets. The patterns of chord changes in AAM seem much more uniform and consistent, but in all datasets there is a clear linear pattern that represents the most likely distance between the currently played chord and the next one. This confirms the existence of popular chord progressions, such as [C:maj, F:maj, G:maj] or [C:maj, G:maj, A:min, F:maj].

\subsection{Data preprocessing} \label{Data preprocessing}

For both model architectures, the data was preprocessed in the same way as
in the papers introducing them.

For BTC, each 10-second audio signal (with consecutive signals overlapping by 5 seconds) was processed at a sampling rate of 22,050 Hz using Constant-Q transform, covering 6 octaves starting from C1, with 24 bins per octave and a hop size of 2048. The CQT features were then converted to log amplitude and global z-normalization was applied, using the mean and variance calculated from the training data. Additionally, pitch augmentation was performed on the audio files with corresponding adjustments to the labels to reflect the pitch changes. Pitch augmentation ranging from -5 to +6 semitones was applied to all training data. 

For HT, for each track from AAM and Winterreise the NNLS chroma features were computed using the Chordino VAMP plugin \cite{NNLSChroma}, installed into the Sonic Visualizer software\cite{SonicVisualiser}. Because this was a highly manual process, for AAM a~subset of 192 out of 3000 songs was 
selected. The AAM tracks, which were saved in 44100Hz, were preprocessed with default settings (window size of 16384, window increment of 2048), resulting in time frames of around 0.046 seconds. However, 
since Winterreise tracks are saved in 22050Hz, they needed to be processed with window size of 8192 and window increment of 1024 to achieve the same result. In both cases, the files were saved to include the timestamp in seconds before the feature, which contains 12 treble chroma and 12 bass chroma. Each input sequence for the Harmony Transformer consists of 100 segments (approximately 23 seconds), created using a sliding window with a frame size of 21 and a hop size of 5. All training data are augmented by shifting the pitch in both the input data and the annotation, expanding the training set 12 times.

Furthermore, annotations for Winterreise came in .csv format, and annotations for AAM in .arff format, so they 
both needed to be converted to .lab using functions created for this purpose.  A couple of the Non-chord labels in AAM tracks are the rows from the annotation files where a value 'BASS NOTE EXCEPTION' was found in the original dataset and converted into the 'N' label. Table~\ref{table_preprocessed_data} summarizes the number of songs and individual training samples in each preprocessed  dataset.

\begin{table}[h]
\caption{Number of songs and training/validation sequences in preprocessed dataset for each model.}
\label{table_preprocessed_data}
\begin{tabular*}{\textwidth}{@{\extracolsep\fill}ccccc}
\toprule
Model & Dataset & Total Songs & Total Instances (Train + Valid) \\
\midrule
\multirow{2}{*}{BTC} 
& AAM & 3000 & 1\,072\,908 \\
& Winterreise & 48 & 18\,312 \\
\midrule
\multirow{3}{*}{HT} 
& AAM & 192 & 13\,766 \\
& Winterreise & 48 & 3\,775 \\
& Billboard & 739 & 73\,550 \\
\bottomrule
\end{tabular*}
\end{table}


%
\section{Experimental setup}

The main focus of the experiments was to check how useful the artificially generated dataset (AAM) is in training chord recognition systems.
A total of 13 experiments (training cycles) were conducted, each with a different training set. Additionally, we present the results of the pre-trained BTC without any additional training as experiment number 0. Each experiment was conducted using six-fold cross-validation. When creating the split into training and validation sets, it was ensured that the two recordings of the same song in the Winterreise dataset always ended up in the same set (either training or validation).

\begin{table}[h]
\caption{Training and evaluation datasets for conducted experiments.}
\label{table_experiments}
\begin{tabular*}{\textwidth}{@{\extracolsep\fill}cccc}
\toprule
Experiment id & Training datasets & Model & Evaluation datasets \\
\midrule

0 & -- & pretrained BTC & \multirow{7}{*}{AAM, Winterreise} \\
\cmidrule(lr){1-3}
1 & AAM & \multirow{3}{*}{BTC} & \\
\cmidrule(lr){1-2} 
2 & Winterreise &                      & \\
\cmidrule(lr){1-2}
3 & AAM, Winterreise &                & \\
\cmidrule(lr){1-3}

4 & AAM & \multirow{3}{*}{pretrained BTC} & \\
\cmidrule(lr){1-2}
5 & Winterreise &                       & \\
\cmidrule(lr){1-2}
6 & AAM, Winterreise &                 & \\
\midrule

7  & Billboard & \multirow{7}{*}{HT} & \multirow{7}{*}{Billboard, AAM, Winterreise} \\
\cmidrule(lr){1-2}
8  & AAM &                       & \\
\cmidrule(lr){1-2}
9  & Winterreise &                & \\
\cmidrule(lr){1-2}
10 & Billboard, AAM &           & \\
\cmidrule(lr){1-2}
11 & Billboard, Winterreise &    & \\
\cmidrule(lr){1-2}
12 & AAM, Winterreise &          & \\
\cmidrule(lr){1-2}
13 & Billboard, AAM, Winterreise & & \\
\bottomrule
\end{tabular*}
\end{table}

In the experiments on BTC, two approaches were tested: (1) training the entire model from scratch and (2) finetuning the  
last fully-connected layer in the pre-trained model provided by the authors. During fine-tuning, weights in all layers except the last one were frozen. Authors of HT do not share the weights of a pretrained model, therefore only the first approach is used in this case.

For all datasets, the annotations with MajMin vocabulary were used, containing 
12 Major chords, 12 Minor chords and the non-chord symbol N. The BTC version for a~standard vocabulary (25 classes) was used in the experiments.

Winterreise consists of 48 audio tracks of 24 different songs, AAM has 3000 tracks and Billboard contains 890 tracks, 739 of which are used when training the HT. 

In the experiments with only one training dataset (1, 2, 3, 7, 8, and 9), the entire available dataset 
was used. In other cases, to create a balanced training dataset, a subset of 192 songs was taken from AAM and Billboard datasets, and each of the 48 songs in the Winterreise dataset was repeated 4 times.

In all experiments, WCSR with Root and MajMin metrics, and Weighted Accuracy with the CCM were used as evaluation metrics.

\section{Results}

\subsection{Experiments on BTC} \label{Experiments on BTC}

\begin{figure}[!h]
  \centering
  \begin{subfigure}[t]{0.5\textwidth}
    \centering
    \includegraphics[width=\textwidth]{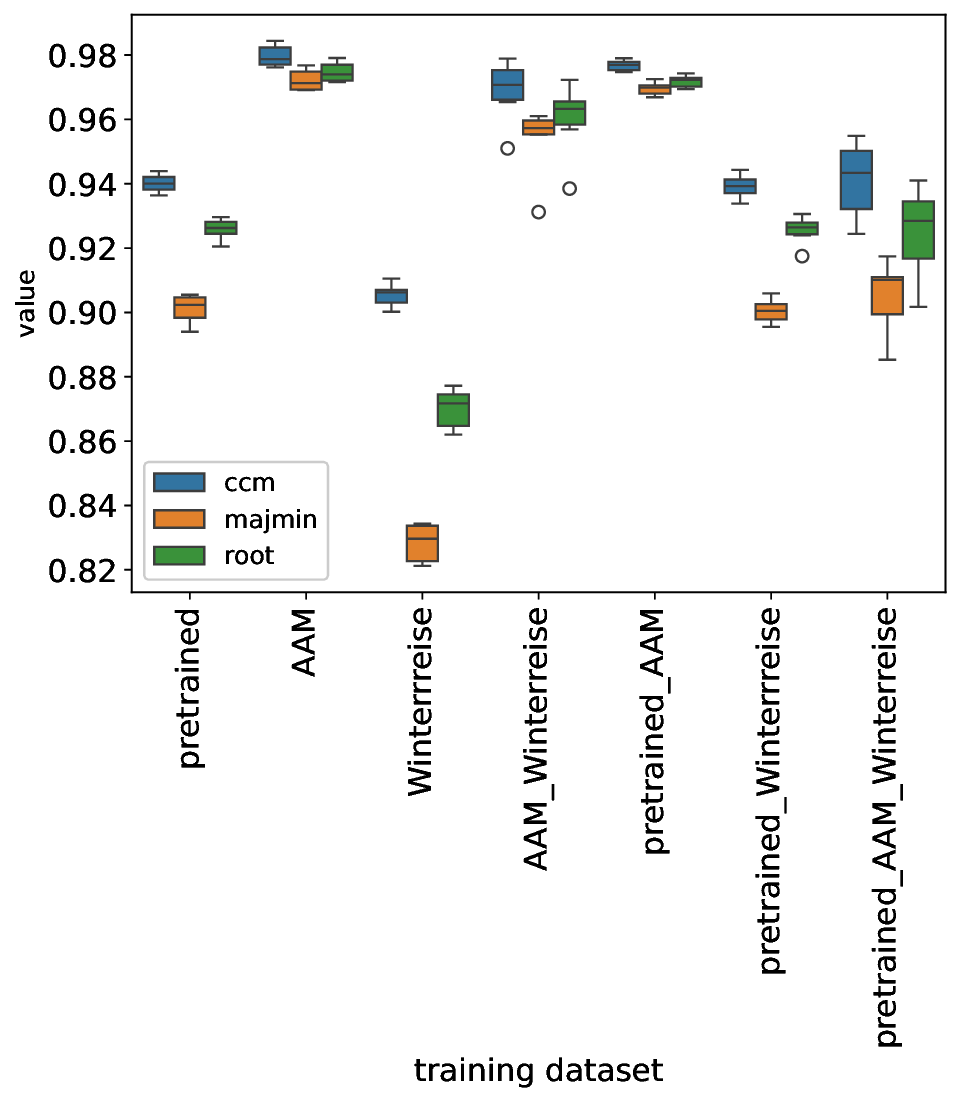}
    \caption{AAM as a validation dataset}
    \label{fig_results_btc_aam}
  \end{subfigure}%
  \begin{subfigure}[t]{0.49\textwidth}
    \centering
    \includegraphics[width=\textwidth]{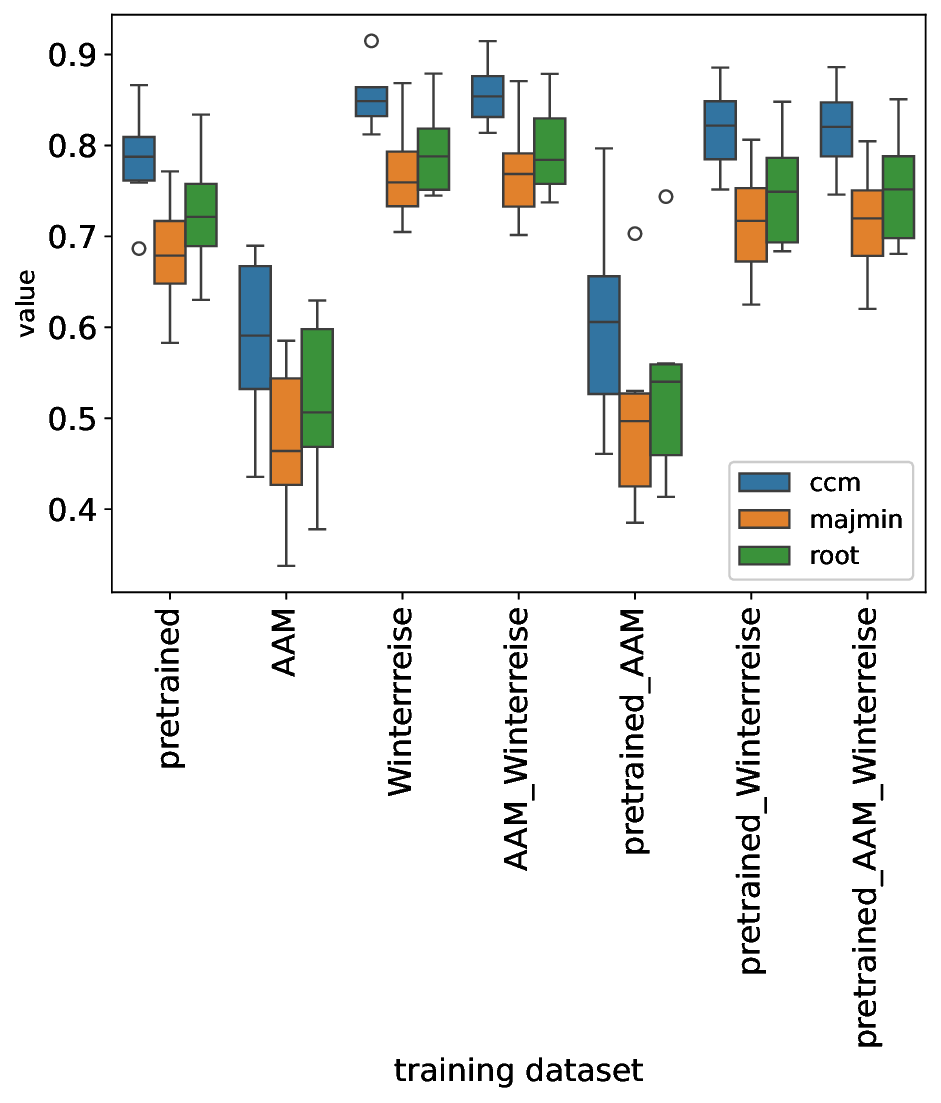}
    \caption{Winterreise as a validation dataset}
    \label{fig_results_btc_winterreise}
  \end{subfigure}
  \caption{Values of evaluation metrics for all experiments on BTC model. Note the different scales on the Y-axes in the charts.}
  \label{fig_results_btc}
\end{figure}

Figure~\ref{fig_results_btc_aam} shows the 
outcomes on the AAM dataset. The results have 
small standard deviations, suggesting that the songs in this dataset have a fairly even complexity 
across all 6 folds. The model trained only on Winterreise achieves the lowest performance for all metrics. The performance of the pre-trained models that were not trained on AAM (pre-trained and pre-trained\_Winterreise) is better, suggesting that a 
significant part of the complexity of the AAM dataset was captured by the corpus of pop songs that BTC was trained on. The pre-trained BTC model fine-tuned on both AAM and Winterreise (pretrained\_AAM\_Winterreise) is performing worse than the one trained on the same corpus from scratch.

Figure~\ref{fig_results_btc_winterreise} shows the values of metrics on the Winterreise dataset. The 
results are, in general, lower than in Figure~\ref{fig_results_btc_aam}, suggesting that this dataset is more difficult to predict than the artificially created one. Furthermore, the results 
have a higher standard deviation, which means that songs in some folds were much more difficult to predict than the others. Additional investigation revealed that songs in fold number 4 (songs with numbers Schubert\_D911-17 to Schubert\_D911-20) seem to be the most challenging. Unsurprisingly, the models that were not trained on Winterreise (AAM and pretrained\_AAM) 
achieved the lowest values for all metrics. For this dataset it does not make a huge difference whether the model was trained from scratch or fine-tuned, which 
suggests that weights learned on the corpus of pop songs that BTC was pre-trained on, actually do not as much help the model to correctly predict chords in Winterreise as they did for AAM.

\subsection{Experiments on HT} \label{Experiments on HT}

\begin{figure}[!h]
  \centering
  \begin{subfigure}{0.5\textwidth}
    \centering
    \includegraphics[width=\textwidth]{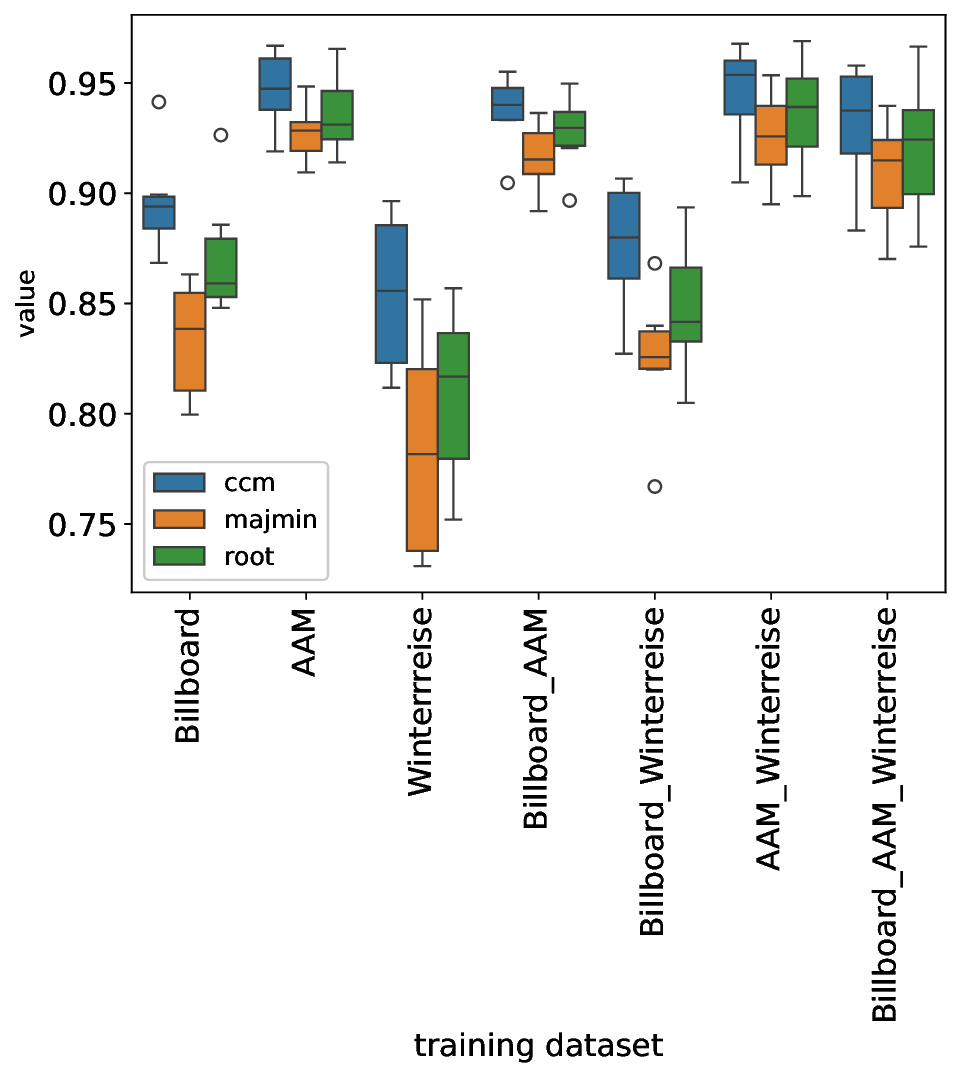}
    \caption{AAM as a validation dataset.}
    \label{fig_results_ht_aam}
  \end{subfigure}%
  \hfill
  \begin{subfigure}{0.5\textwidth}
    \centering
    \includegraphics[width=\textwidth]{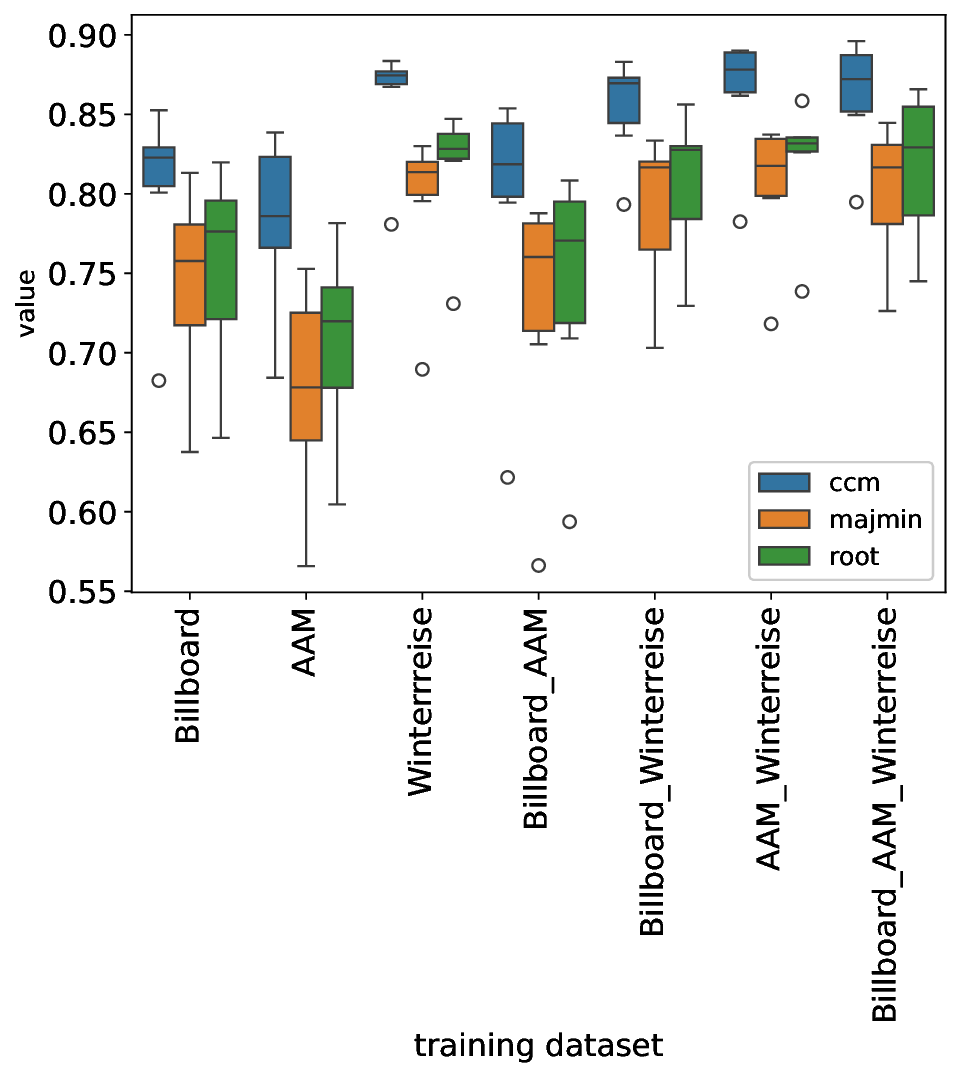}
    \caption{Winterreise as a validation dataset.}
    \label{fig_results_ht_winterreise}
  \end{subfigure}

  \vspace{1em} 

  \begin{subfigure}{0.5\textwidth}
    \centering
    \includegraphics[width=\textwidth]{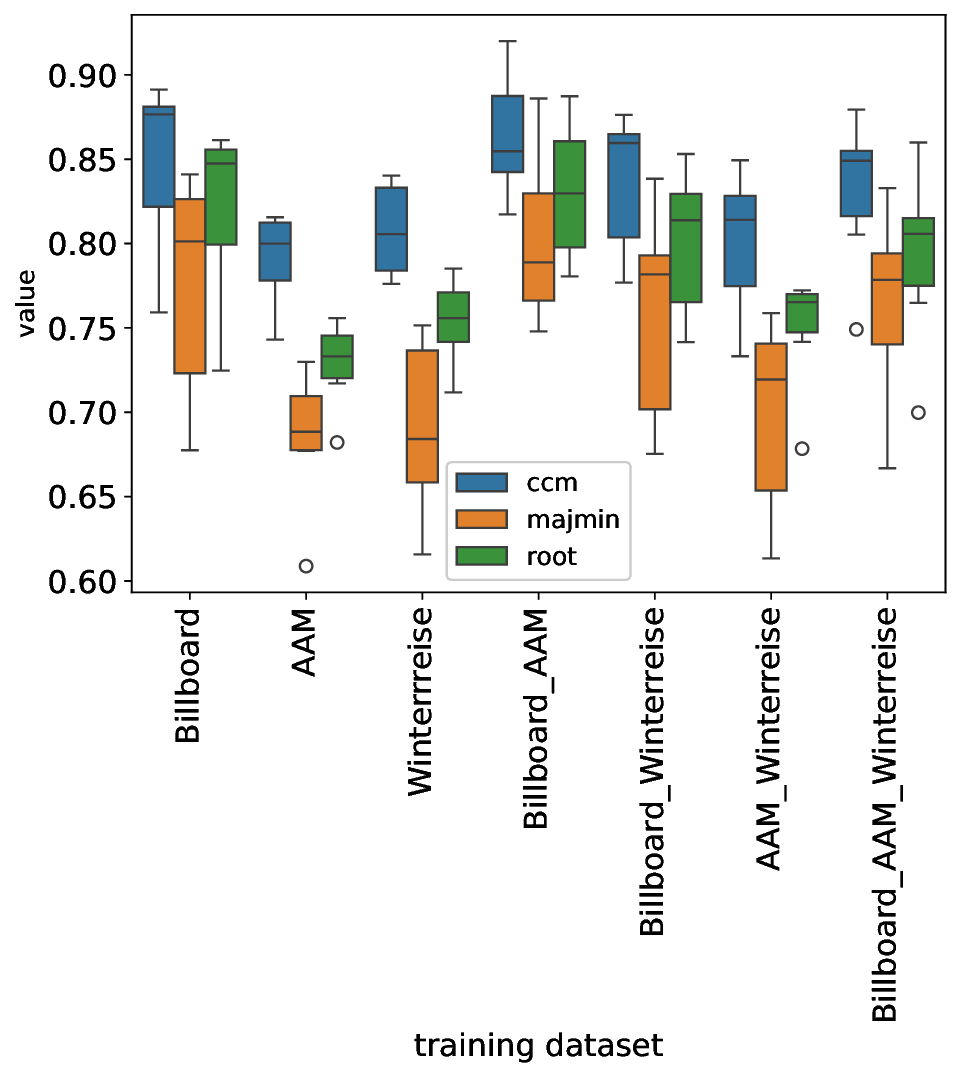}
    \caption{Billboard as a validation dataset.}
    \label{fig_results_ht_billboard}
  \end{subfigure}
  \caption{Values of evaluation metrics for all experiments on HT model. Note the different scales on the Y-axes in the charts.}
  \label{fig_results_ht}
\end{figure}

Figure~\ref{fig_results_ht_aam} shows the values of metrics on the AAM dataset. The performance of the model trained from scratch on Billboard is better than the one trained on Winterreise, suggesting that Billboard and AAM are more similar to each other than Winterreise and AAM.

The results on Winterreise for all experiments 
with the HT model are presented in Figure~\ref{fig_results_ht_winterreise}. The outliers visible in the figure are again 
linked to fold number 4. The model trained on Billboard is performing slightly better than the one trained on AAM. Adding AAM to the training dataset does not seem to improve performance (comparing Billbaord to Billboard\_AAM, Winterreise to AAM\_Winterrise and Billboard\_Winterreise to Billboard\_AAM\_Winterreise).

Figure~\ref{fig_results_ht_billboard} shows the values of metrics on the Billboard dataset. The results of a model trained on AAM are comparable to those of the model trained on Winterreise and the one trained on both datasets, which suggests that AAM 
is a decent training set for a model that aims to predict chord sequences in pop music, if no other data is available. In this case, adding a 
large artificial corpus to a small set of classical music (comparing Winterreise to AAM\_Winterreise) improves performance. 
This observation indicates that in terms of chord sequence similarity, AAM and Billboard are 
closer to each other than AAM and Winterreise.

\subsection{Statistical significance of results.}

The mean and standard deviation of metric values presented on the plots in Section~\ref{Experiments on BTC} are summarized in Table~\ref{table_results} (experiments 0-6). Again, it is visible that results on Winterreise have lower means and higher standard deviations. The highest evaluation values on AAM were achieved by a model trained from scratch on AAM (experiment~1) and the highest evaluation values on Winterreise were achieved by a model trained from scratch on both AAM and Winterreise (experiment~3). However, the difference between the latter and the one trained just on Winterreise (experiment~2) is not statistically significant.

The mean and standard deviation of metric values presented in Section~\ref{Experiments on HT} are summarized in Table~\ref{table_results} (experiments 7-13). For predicting chord sequences in AAM, the best models were these trained on AAM (experiment~8) or on AAM\_Winterreise (experiment~12), depending on the evaluation metric. In terms of predicting Winterreise, the top model was AAM\_Winterreise model (experiment~12), and for Billboard - the Billboard\_AMM model (experiment~10).

\begin{sidewaystable}
\caption{Mean and standard deviation from results of 6 fold validation from experiments on BTC and HT model. The largest mean value in each column is bolded.}
\label{table_results}
\begin{tabular}{lccc|ccc|ccc}
\toprule
& \multicolumn{3}{@{}c@{}}{AAM} & \multicolumn{3}{@{}c@{}}{Winterreise} & \multicolumn{3}{@{}c@{}}{Billboard} \\
\cmidrule{2-4} \cmidrule{5-7} \cmidrule{8-10}
Exp id & Root & MajMin & CCM & Root & MajMin & CCM & Root & MajMin & CCM \\
\midrule
0 & 92.59 $\pm$ 0.33 & 90.11 $\pm$ 0.46 & 94.01 $\pm$ 0.29 & 72.59 $\pm$ 7.05 & 67.99 $\pm$ 6.61 & 78.29 $\pm$ 6.06 & - & - & - \\
1 & \textbf{97.47} $\pm$ 0.32 & \textbf{97.22} $\pm$ 0.34 & \textbf{97.97} $\pm$ 0.35 & 51.74 $\pm$ 9.69 & 47.25 $\pm$ 9.26 & 58.54 $\pm$ 9.88 & - & - & - \\
2 & 87.01 $\pm$ 0.63 & 82.84 $\pm$ 0.62 & 90.54 $\pm$ 0.37 & 79.46 $\pm$ 5.26 & 77.02 $\pm$ 5.92 & 85.35 $\pm$ 3.63 & - & - & - \\
3 & 96.01 $\pm$ 1.17 & 95.36 $\pm$ 1.12 & 96.88 $\pm$ 1.00 & \textbf{79.64} $\pm$ 5.46 & \textbf{77.15} $\pm$ 5.99 & \textbf{85.74} $\pm$ 3.70 & - & - & - \\
4 & 97.18 $\pm$ 0.20 & 96.96 $\pm$ 0.21 & 97.68 $\pm$ 0.17 & 53.92 $\pm$ 11.73 & 50.33 $\pm$ 11.37 & 60.71 $\pm$ 11.99 & - & - & - \\
5 & 92.55 $\pm$ 0.45 & 90.04 $\pm$ 0.38 & 93.92 $\pm$ 0.38 & 75.06 $\pm$ 6.58 & 71.47 $\pm$ 6.64 & 81.85 $\pm$ 4.99 & - & - & - \\
6 & 92.48 $\pm$ 1.47 & 90.50 $\pm$ 1.20 & 94.12 $\pm$ 1.22 & 75.26 $\pm$ 6.58 & 71.50 $\pm$ 6.58 & 81.77 $\pm$ 5.08 & - & - & - \\
\midrule
7 & 
87.16 $\pm$ 3.00 & 
83.35 $\pm$ 2.74 & 
89.64 $\pm$ 2.48 & 
75.40 $\pm$ 6.53 & 
74.31 $\pm$ 6.28 & 
80.18 $\pm$ 6.09 & 
82.04 $\pm$ 5.47 & 
77.55 $\pm$ 7.01 & 
84.83 $\pm$ 5.39 \\
8 & 
93.59 $\pm$ 1.90 & 
\textbf{92.74} $\pm$ 1.35 & 
\textbf{94.68} $\pm$ 1.83 & 
70.64 $\pm$ 6.26 & 
67.50 $\pm$ 6.85 & 
78.17 $\pm$ 5.65 & 
72.83 $\pm$ 2.64 & 
68.44 $\pm$ 4.22 & 
79.10 $\pm$ 2.84 \\
9 & 
80.88 $\pm$ 4.10 & 
78.38 $\pm$ 5.20 & 
85.45 $\pm$ 3.69 & 
81.59 $\pm$ 4.27 & 
79.40 $\pm$ 5.24 & 
85.98 $\pm$ 3.90 & 
75.36 $\pm$ 2.63 & 
68.99 $\pm$ 5.40 & 
80.78 $\pm$ 2.84 \\
10 & 
92.73 $\pm$ 1.82 & 
91.61 $\pm$ 1.61 & 
93.69 $\pm$ 1.80 & 
74.13 $\pm$ 8.13 & 
72.69 $\pm$ 8.50 & 
79.28 $\pm$ 8.69 & 
\textbf{83.09} $\pm$ 4.22 & 
\textbf{80.23} $\pm$ 5.23 & 
\textbf{86.39} $\pm$ 3.83 \\
11 & 
84.77 $\pm$ 3.15 & 
82.44 $\pm$ 3.32 & 
87.60 $\pm$ 3.06 & 
80.69 $\pm$ 4.74 & 
78.98 $\pm$ 5.22 & 
85.43 $\pm$ 3.38 & 
80.10 $\pm$ 4.52 & 
75.82 $\pm$ 6.67 & 
83.73 $\pm$ 4.39 \\
12 & 
\textbf{93.62} $\pm$ 2.56 & 
92.55 $\pm$ 2.14 & 
94.54 $\pm$ 2.35 & 
\textbf{82.04} $\pm$ 4.17 & 
\textbf{80.39} $\pm$ 4.54 & 
\textbf{86.34} $\pm$ 4.13 & 
74.90 $\pm$ 3.63 & 
69.84 $\pm$ 6.00 & 
80.09 $\pm$ 4.38 \\
13 & 
92.07 $\pm$ 3.27 & 
90.88 $\pm$ 2.58 & 
93.09 $\pm$ 2.87 & 
81.74 $\pm$ 4.85 & 
80.15 $\pm$ 4.47 & 
86.21 $\pm$ 3.76 & 
79.24 $\pm$ 5.46 & 
76.39 $\pm$ 5.83 & 
83.15 $\pm$ 4.70 \\
\bottomrule
\end{tabular}
\end{sidewaystable}

%
\subsection{Computational complexity.}

All experiments were conducted using the resources of an HPC (High Performance Computing) cluster on a single GPU.
Training times varied according to the experiment and the fold number. 
Table~\ref{table_times} shows the number of epochs needed to complete training for each fold and the average number of minutes a single epoch lasted. The longest experiments, 1st and 4th, which used the whole AAM dataset, took up to 10 days to compute.

\begin{table}[h]
\caption{Number of epochs taken to complete each of the 6 folds for all experiments and estimated time one epoch took to complete (in minutes)}\label{table_times}
\centering
\begin{tabular}{ccccc}
\toprule
\textbf{Exp id} & \textbf{Model} & \textbf{Training datasets} & \textbf{Epochs} & \textbf{Time} \\
\midrule
1 & \multirow{3}{*}{BTC} & AAM & 48, 65, 52, 47, 64, 54 & 150 \\
2 &  & Winterreise & 36, 21, 18, 18, 56, 20 & 2 \\
3 &  & AAM, Winterreise & 35, 19, 35,  15, 37, 27  &  13 \\
4 & \multirow{3}{*}{pretrained BTC} & AAM & 79, 64, 42, 29, 50, 71 &  180 \\
5 &  & Winterreise & 19, 32, 11, 24, 3, 16 & 2 \\
6 &  & AAM, Winterreise & 44, 22, 15, 41, 1, 4 & 11 \\
7 & \multirow{7}{*}{HT} & Billboard & 27, 17, 24, 18, 15, 27 & 6 \\
8 &  & AAM &  20, 55, 26, 5, 34, 16 & 1 \\
9 &  & Winterreise & 35, 16, 22, 57, 33, 28 & 1 \\
10 &  & Billboard, AAM &  22, 10, 28, 20, 31, 19 & 3 \\
11 &  & Billboard, Winterreise &  4, 23, 28, 47, 16, 33 & 4 \\
12 &  & AAM, Winterreise & 68, 40, 33, 34, 33, 19 & 3 \\
13 &  & Billboard, AAM, Winterreise &  9, 23, 11, 25, 27, 26 & 4 \\
\bottomrule
\end{tabular}
\end{table}

\subsection{Summary of results}

Conducted experiments suggest that for both model architectures, the chord progressions are easiest to predict in the AAM dataset, with Billboard being more challenging and then Winterreise being the hardest. This might be due to the fact that both Billboard and Winterreise have annotations with a larger, more complicated vocabulary, and these experiments use a simplified MinMaj one. For AAM, only a MajMin vocabulary is present, so the segment labeled as a certain chord is matched exactly,without any additional notes.

For the Bidirectional Transformer for Chord Recognition, training from scratch generally worked better than fine-tuning a pretrained model in the scenario where the training and evaluation datasets were the same. However, starting from pretrained weights improved the evaluation of the model on the AAM dataset when trained on Winterreise.

For the Harmony Transformer, enriching the training dataset with AAM improved performance for both Winterreise and Billboard by about 1 percentage point. BTC and HT, as neural network models for Audio Chord Recognition based on the Transformer architectures, can achieve comparable results, depending on the training and validation datasets.

\section{Conclusions and future work}

The experiments conducted in this work prove that 
artificially generated music datasets such as Artificial Audio Multitracks can be useful in certain scenarios. For instance, in enriching a smaller training dataset of music composed by a human or even as a training set for a model which purpose is to predict chord sequences in pop music, if no other data is available.

In future works, one could consider automating the process of creating NNLS chroma features using the Chordino plugin and processing the entire AAM dataset instead of a subset, and repeating the experiments using HT. Another possible improvement would be recreating the dataset used for training BTC by downloading songs from streaming services and including them in similar experiments. One could also consider conducting similar experiments with a larger vocabulary if an artificially generated dataset with richer annotations is released.

\section*{Acknowledgements}
This research was carried out with the support of the Laboratory of Bioinformatics and Computational Genomics and the High Performance Computing Center of the Faculty of Mathematics and Information Science Warsaw University of Technology.

\bibliographystyle{unsrt}  
\bibliography{references}  

\end{document}